\documentclass{article}

\usepackage{arxiv}

\usepackage[utf8]{inputenc} % allow utf-8 input
\usepackage[T1]{fontenc}    % use 8-bit T1 fonts
\usepackage{hyperref}       % hyperlinks
\usepackage{url}            % simple URL typesetting
\usepackage{booktabs}       % professional-quality tables
\usepackage{amsfonts}       % blackboard math symbols
\usepackage{nicefrac}       % compact symbols for 1/2, etc.
\usepackage{microtype}      % microtypography
\usepackage{lipsum}
\usepackage{xspace}
\usepackage{amsmath,amssymb}
\usepackage{graphicx}
\usepackage{cuted}
\usepackage{multicol}

\usepackage[symbol]{footmisc}

\usepackage[square,comma,sort&compress,numbers]{natbib}
\newcommand{\dbd}[1]{\frac{d}{d #1}}
\newcommand{\wrt}[1]{\text{d} #1}
\newcommand*{\ie}{\emph{i.e.}\@\xspace}

\newcommand{\figref}[2][]{\mbox{Fig.~\ref{fig:#2}#1}}

\renewcommand\citep{\cite}
\renewcommand\citet{\cite}

\title{Firing rate of the leaky integrate-and-fire neuron with stochastic conductance-based synaptic inputs with short decay times}

\author{
 \makebox[.4\linewidth]{Timothy D. Oleskiw}\\
  Center for Neural Science\\
  New York University\\
  New York City, NY, USA\\
\And
 \makebox[.4\linewidth]{Wyeth Bair*}\\
  Department of Biological Structure\\
  University of Washington\\
  Seattle, WA, USA\\
\And
 \makebox[.4\linewidth]{Eric Shea-Brown*}\\
  Department of Applied Mathematics\\
  University of Washington\\
  Seattle, WA, USA\\
\And
\makebox[.4\linewidth]{Nicolas Brunel}\\
  Departments of Neurobiology and Physics\\
  Duke University\\
  Durham, NC, USA\\
}

\begin{document}

\twocolumn[\maketitle
\begin{abstract}
We compute the firing rate of a leaky integrate-and-fire (LIF) neuron with stochastic conductance-based inputs in the limit when synaptic decay times are much shorter than the membrane time constant.  A comparison of our analytical results to numeric simulations is presented for a range of biophysically-realistic parameters.
\end{abstract}\vspace{5pt}]

\section*{Introduction}
\label{introduction}
\footnotetext{*These authors contributed equally to this work.}
Information processing within neural networks is widely considered to be achieved by circuit computations in which the firing rate, either of a single neuron or populations of functionally similar neurons, serves as the fundamental variable \citep{barlow72perception}.  Therefore, by understanding how basic mathematical operations like addition and multiplication are applied to firing rates in networks, we may gain insight into fundamental mechanisms of neural computation \citep{carandini12normalization, yu02softmax,angelucci06connections,sato14distal}.

\par
Multiple studies have demonstrated that a neuron's output rate can be significantly affected by the \emph{timescale} of fluctuating input, which can be modulated by factors such as the spike timing and correlation of upstream activity \citep{moreno02correlated} or by the kinetics of synaptic filtering \citep{odonnell14voltage}.  Input timescale has been shown in several studies to impact the firing rates of model neurons \citep{brunel98lif,fourcaud02lif,moreno02correlated,mitchel03shunting, brunel03quadratic} as well as the gain and phase of their frequency response \citep{brunel01noise,fourcaud02lif}.

\par
Most of our present insights about how the time scales of synaptic inputs affect output firing rate come from analytic solutions for the firing rate of leaky integrate-and-fire (LIF) neurons under stochastic input, in both the short \cite{brunel98lif, fourcaud02lif} and long input time limits \cite{moreno02correlated}. 
Recently, the firing rate of a LIF neuron for arbitrary input time scale was obtained as a solution of a Fredholm integral equation of the second kind, which can then be solved numerically \cite{vanvreeswijk19fredholm}. However, these studies all use \emph{current based} LIFs, \ie, synaptic input to be injected current.  Such a formulation has the advantage of simplicity and may be valid in some physiological limits, but neglects the general dependence of synaptic inputs on membrane potential.  These dependencies can be described by the so-called `conductance-based' formalism wherein synaptic inputs are a product of synaptic conductance times the `driving force', \ie the difference between membrane potential and the synaptic reversal potential.

\par
Here, we present a generalization of the calculations described in \cite{brunel98lif, fourcaud02lif} to a LIF model with  conductance-based synaptic inputs, when the input correlation time constant is much shorter than the membrane time constant.

\section*{Model formulation}
\vspace{-5pt}
A ubiquitous model of a one-compartmental neuron with conductance-based synaptic inputs is defined by its membrane potential $V(t)$, and excitatory (E) and inhibitory (I) total conductances $g_e(t)$ and $g_i(t)$, whose dynamics obey
\begin{eqnarray}
	C \dbd{t}V & = & g_l(v_l-V) + g_e(v_e-V) \\ \nonumber
	& \quad & + g_i(v_i-V)\label{eq:v}\\
	\tau_e \dbd{t}g_e & = & -g_e + \mu_e + \sigma_e \sqrt{\tau_e} \eta_e(t)\label{eq:ge}\\
	\tau_i \dbd{t}g_i & = & -g_i + \mu_i +\sigma_i \sqrt{\tau_i}\eta_i(t)\label{eq:gi},
\end{eqnarray}
where $C$ is the membrane capacitance, $g_l$ the leak conductance, $v_l$ the resting membrane potential, $v_e$ and $v_i$ are synaptic reversal potentials, $\tau_e$ and $\tau_i$ are synaptic decay time constants, $\mu_e$ and $\mu_i$ are mean synaptic conductances, $\sigma_e$ and $\sigma_i$ are the amplitude of the fluctuations, and $\eta_e$ and $\eta_i$ Gaussian white noise, each for the E and I conductances, respectively \cite{ayaz09modulation,ly09dynamic}. These equations are complemented with the usual threshold-and-reset mechanism, \ie a spike is emitted whenever the voltage reaches a threshold $v_t$ and the voltage is then reset instantaneously to $v_r$.  Equations \eqref{eq:ge} and \eqref{eq:gi} are obtained from Poisson synaptic inputs using a diffusion approximation, \ie a shot noise process approximated by a continuous Gaussian process with the same mean and variance.  As the post-synaptic potentials evoked by neural inputs are not instantaneous, timescales $\tau_e$ and $\tau_i$ are chosen to mimic excitatory and inhibitory neurotransmitter kinetics.

Equations \eqref{eq:v}-\eqref{eq:gi} are difficult to analyze mathematically because the computation of quantities of interest (mean firing rate, distribution of membrane potential) involves solving a 3D Fokker-Planck equation with complicated boundary conditions at threshold and reset. A first simplification consists in considering that only a single type of conductance fluctuates (here, E), while the other is constant in time, $g_i=\mu_i$, leading to the two-variable system
\begin{align}\label{eq:gs}
\begin{split}
	C \dbd{t}V & = g_l(v_l-V) + g_e(v_e-V) +\mu_i (v_i-V) \\
	\tau_e \dbd{t}g_e & = -g_e + \mu_e + \sigma_e \sqrt{\tau_e} \eta(t). 
\end{split}
\end{align}

Rewriting $g_e = \mu_e+\sigma_e z$, and $\tau_e=\tau_s$, we obtain
\begin{align}\label{eq:general_lif}
\begin{split}
\tau_m \dbd{t}V & =  A(V) + \frac{1}{k} B(V)z \\
\tau_s \dbd{t}z & = -z + \sqrt{\tau_s} \eta(t),
\end{split}
\end{align}
where the membrane time constant $\tau_m=C/g_l$, $k=\sqrt{\tau_s/\tau_m}$, and
\begin{align}
\begin{split}\label{eq:ab}
A(V) &= v_l-V+ \frac{\mu_e}{g_l}(v_e-V) + \frac{\mu_i}{g_l} (v_i-V) \\
B(V) &= \frac{\tilde \sigma_e}{g_l}  (v_e -V),
\end{split}
\end{align}
where $\tilde \sigma_e= k \sigma_e$. Note that $\tilde \sigma_e$ should be of order 1 in the limit $k\rightarrow 0$ for conductance fluctuations to lead to fluctuations of the voltage of finite variance.  This means that $\sigma_e$ should be of order $1/k$ in that limit. We now seek to approximate the firing rate $r$ of system \eqref{eq:general_lif} by solving for the mean number of threshold crossings (spikes) per unit time (seconds) under general input conditions.

\section*{Simulation methods}
\label{methods}

\par
Simulations of all spiking LIF models were performed in MATLAB R2013b.  Dynamics were evaluated numerically with the forward Euler method at a time step of 10 microseconds.  After crossing threshold, a spike was recorded and membrane voltage was forced to reset instantaneously.  Spike-rate response was determined from the mean spike frequency over a 100 second stimulation duration.  Code is available upon request.

\section*{Results}
\label{results}

\par
We now demonstrate the key steps to approximating the firing rate of the general LIF system described by \eqref{eq:general_lif}.  The associated equilibrium Fokker-Planck equation for the distribution $P(V,z)$ of voltage $V$ and input $z$ is given by \cite{gardiner09handbook}
\begin{equation}\label{eq:fp}
\mathcal{L} P - kz \frac{\partial}{\partial V}(B(V)P) -k^2 \frac{\partial}{\partial V} (A(V)P) = 0,
\end{equation}
with the differential operator $\mathcal{L}$ defined as
\begin{equation}
\mathcal{L} P = \frac{1}{2} \frac{\partial^2 P}{\partial z^2} +\frac{\partial}{\partial z}(zP). 
\end{equation}
The probability flux in voltage $V$ is therefore
\begin{equation}
J_V = \frac{1}{\tau_m} \left(A(V) + B(V) \frac{z}{k}\right) P,
\end{equation}
which cannot be negative at spiking threshold $V=V_{th}$, giving rise to the boundary conditions
\begin{align}
\begin{split}
P(V_{th},z) = 0 &  \quad z< -k \frac{A(V_{th})}{B(V_{th})} \\
P(V_{th},z) \ge 0 & \quad z> -k \frac{A(V_{th})}{B(V_{th})}.
\end{split}
\end{align}
The strategy is to find solutions in boundary layers, as in \cite{hagan89halfline} and \cite{fourcaud02lif}.  We compute the solution in three regions: $(i)$ in the \emph{outer} region far from both threshold and reset, $(ii)$ in the \emph{threshold layer} when $V$ is close to emitting a spike, and $(iii)$ in the \emph{reset layer} when $V$ is close to the reset potential.

\subsection*{Outer solution}

The outer solution, far from reset and threshold, is obtained by expanding the probability distribution $P$ in powers of $k$, \ie $P=P_0+ k P_1 + k^2 P_2 +\ldots$.  Substituting this expansion into \eqref{eq:fp}, we find a recurrence relation for the distribution terms $P_i$ given by
\begin{align}
\begin{split}
\mathcal{L} P_0 & = 0 \\
\mathcal{L} P_1 & = z \frac{\partial}{\partial V}(BP_0) \\
\mathcal{L} P_2 & = z \frac{\partial}{\partial V}(BP_1)+ \frac{\partial}{\partial V} (A P_0) \\
& \ldots, 
\end{split}
\end{align}
which leads to
\begin{align}
\begin{split}
P_0 & = \frac{e^{-z^2}}{\sqrt{\pi}} Q_0(V) \\
P_1 & =  \frac{e^{-z^2}}{\sqrt{\pi}} Q_1(V) -\frac{z e^{-z^2}}{\sqrt{\pi}} \frac{\partial}{\partial V} (BQ_0) \\
P_2 & =  \frac{e^{-z^2}}{\sqrt{\pi}} Q_2(V) -\frac{z e^{-z^2}}{\sqrt{\pi}} \frac{\partial}{\partial V} (BQ_1) \\
 & \quad + \frac{z^2 e^{-z^2}}{2\sqrt{\pi}} \frac{\partial}{\partial V} \left(B \frac{\partial}{\partial V} (BQ_0)\right).
\end{split}
\end{align}
To find a solution for $P_2$ that satisfies the boundary conditions on $z$ (\ie, both $P_2$ and $\partial P_2/\partial z$ should go to zero in both $z\rightarrow \pm \infty$ limits) we need to impose the solvability condition
\begin{equation}\label{eq:q0}
\frac{1}{2} \frac{\partial}{\partial V} \left(B(V) \frac{\partial (B(V) Q_0)}{\partial V}\right) - \frac{\partial (A(V) Q_0)}{\partial V} = 0,
\end{equation}
which is solved for $Q_0$. As expected, \eqref{eq:q0} coincides with the Fokker-Plank equation in the white noise limit using Stratonovich calculus. Going to third order, we find that $Q_1$ obeys \eqref{eq:q0} as well. Thus, 
\begin{align}
\begin{split}
Q_0(V) & = \left\{\begin{array}{ll} \alpha_0 R(V) &  V<V_{re} \\ \beta_0 R(V) + \gamma_0 S(V) & V>V_{re}\end{array}\right. \\
Q_1(V) & = \left\{\begin{array}{ll} \alpha_1 R(V) & V<V_{re} \\ \beta_1 R(V) + \gamma_1 S(V) & V>V_{re} \end{array}\right. 
\end{split}
\end{align}
for the voltage reset potential $V_{re}$, where 
\begin{equation}
R(V) = \frac{W(V)}{B(V)}
\end{equation}
and
\begin{equation}
S(V) = \frac{W(V)}{B(V)} \int^{V^t}_{V} \frac{ \wrt{u}}{B(u)W(u)}
\end{equation}
for $W$ given in \eqref{eq:w}.  Furthermore, the solutions $Q_0$ and $Q_1$ have to obey the normalization conditions
\begin{align}
\begin{split}
\int Q_0(V) dV & = 1 \\
\int Q_1(V) dV & = 0.
\end{split}
\end{align}

\subsection*{Inner solutions}

\par
Solutions to the inner threshold and reset layers are found using similar techniques.  To construct solutions within the threshold layer, we need to transform voltage as $V=V_{th} - kxV_{th}$.  It will also be convenient to define a substitution
\begin{equation}
z' = z+ k \frac{A(V_{th})}{B(V_{th})},
\end{equation}
simplifying the boundary conditions and operator $\mathcal{L}'$.  Therefore, rewriting \eqref{eq:fp} for the threshold distribution $P^T$, \ie the probability of the system near spiking threshold $V_th$, we have
\begin{align}
\begin{split}
\mathcal{L}' P^T & + z' \frac{\partial P^T}{\partial x} \\
 & \quad - k \left( \frac{A(V_{th})}{B(V_{th})} \frac{\partial P^T}{\partial z'} + \frac{B(V_{th})}{B(V_{th})} z' \frac{\partial}{\partial x} (x P^T)\right) \\
 &\quad + O(k^2) = 0,
\end{split}
\end{align}
which again can be solved by expanding $P^T=P_o^T +k P_1^T + \ldots$, satisfying the boundary condition $P^T(0,z')=0$ for $z'<0$.  The probability flux at $(0,z')$ is given by $z'P^T(0,z') B(V_{th})/(\tau_m k)$, implying $P_0^T=0$. The firing rate terms at zero and first orders are therefore
\begin{align}
\begin{split}\label{eq:nu1}
\nu_0 & = \frac{B(V_{th})}{\tau_m}\int_0^{\infty} z' P_1^T(0,z')  \wrt{z'}\\
\nu_1 & = \frac{B(V_{th})}{\tau_m}\int_0^{\infty} z' P_2^T(0,z')  \wrt{z'}.
\end{split}
\end{align}
A solution to $P_1^T$ has previously been found by \cite{hagan89halfline} to be
\begin{equation}\label{eq:pt1}
P_1^T  =  \frac{e^{-z'^2}}{\sqrt{\pi}} \rho_1^T \left(\tilde \alpha +x+z + U(x,z)\right),
\end{equation}
where $\tilde \alpha$ and $U(x,z)$ are described in \cite{hagan89halfline}. 
Using the fact that $U$ decays exponentially to zero for large $x$, and that $\int z e^{-z'^2} U(x,z)=0$, we conclude that $\rho_1^T = 2 \nu_0\tau_m/B(V_{th})$ \citep{hagan89halfline}. 

\par
The reset layer can be dealt with exactly in the same way as the threshold layer.  One finds the solutions to the left and right of the reset, \ie $V_{re}^-$ and $V_{re}^+$,
coincide at zero order, but that the difference between these solutions obeys \eqref{eq:pt1} \cite{fourcaud02lif,brunel01noise}.

\subsection*{Matching outer and inner layers}

\par
To match the outer and threshold layers we use the change of variables
$V=V_{th}- k x B(V_{th})$, $z' = z+ k A(V_{th})/B(V_{th})$.
Therefore, the outer solution becomes
\begin{align}
\begin{split}
P(x,z') & = \frac{e^{-z'^2}}{\sqrt{\pi}} \Biggl(Q_0(V_{th}) + k \biggl[Q_1(V_{th})\biggr.\Biggr.\\
& \quad \Biggl.\biggl. -x B(V_{th})Q'_0(V_{th}) + \frac{z'\gamma_0}{B(V_{th})}\biggr]\Biggr).
\end{split}
\end{align}
This solution must match $P_1^T$ in the large $x$ limit.  Hence, we have $Q_0(V_{th})=0$ which implies $\beta_0=0$.  We also have 
\begin{align}
\begin{split}
\gamma_0 & = 2\nu_0 \tau_m \\
Q_1(V_{th}) & = \tilde \alpha \rho_1^T = \frac{2\tilde\alpha\nu_0\tau_m}{B(V_{th})},
\end{split}
\end{align}
which leads to
\begin{equation}
\beta_1 = \frac{2\tilde\alpha\nu_0\tau_m}{W(V_{th})}.
\end{equation}

\par
Matching of the reset and outer layers is done in a similar way. One finds that $Q_0$ has to be continuous in $V_{re}$, implying
\begin{equation}
\alpha_0 = \int^{V_{th}}_{V_{re}}\frac{\wrt{u}}{B(u)W(u)},
\end{equation}
which, together with the normalization condition for $Q_0$, leads to the equation for the zeroth order firing rate $\nu_0$ as expected.  One then finds that $Q_1$ is instead discontinuous in $V_{re}$, with
\begin{align}
\begin{split}
Q_1(V_{re}^+) & - Q_1(V_{re}^-) = \tilde \rho_1^T \\
& = \beta_1 R(V_{re}) +\gamma_1 S(V_{re}) - \alpha_1 R(V_{re}).
\end{split}
\end{align}
Including now the normalization condition for $Q_1$, this gives us two equations for the last two remaining unknowns, $\gamma_1$ and $\alpha_1$. In particular, we find that
\begin{equation}
\gamma_1 =  4 \tilde\alpha (\nu_0\tau_m)^2 B(V_{th})\left(\frac{B(V_{re})^2}{B(V_{th})^2} \psi(V_{re}) - \psi(V_{th})\right),
\end{equation}
where
\begin{equation}
\psi(V) = \frac{1}{B(V)W(V)} \int^V_{-\infty} R(u)du.
\end{equation}

\begin{figure}[t!]\centering
    \vspace{-5pt}
	\includegraphics[scale=.9]{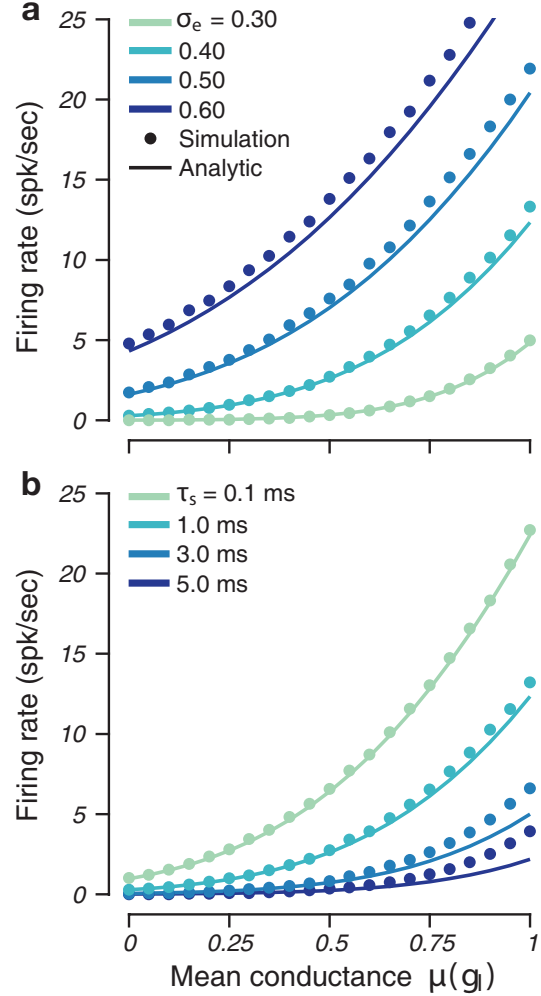}
	\caption{\label{fig:comparison}Comparison of the analytic approximation to the firing rate of a neuron described by \eqref{eq:general_lif}, \eqref{eq:ab}, and \eqref{eq:general_lif_sol} to values estimated through numerical simulation.  (a) Standard deviation of excitatory conductance $\sigma_e$ is varied for fixed timescale $\tau_s = 1$ ms, and (b) synaptic timescale is varied for fixed $\sigma_e = 0.40 g_l$.}
	\vspace{-5pt}
\end{figure}

\subsection*{First order correction to the firing rate}
\par
The last step is to compute $\nu_1$. From \eqref{eq:nu1} it would seem that we need to compute $P_2^T$. Fortunately we only need the term proportional to $z$ in this equation, as it is the only term that contributes to the firing rate.  Further, the condition matching the outer and inner solutions requires this term to be proportional to $\gamma_1$. Therefore, the correction is
\begin{align}
\begin{split}
\label{eq:derivation_last}
\nu_1 & = \frac{\gamma_1}{2\tau_m} \\
 & = 2 \tilde\alpha \nu_0^2 \tau_m B(V_{th})\left(\frac{B(V_{re})^2}{B(V_{th})^2} \psi(V_{re}) - \psi(V_{th})\right).
\end{split}
\end{align}
As in \cite{brunel98lif} and \cite{fourcaud02lif} we can express the firing rate $r$ as
\begin{equation}\label{eq:general_lif_sol}
	\frac{1}{r} = 2\tau_m\int^{v_{th}^{eff}}_{v_{re}^{eff}}\frac{\wrt{z}}{B(z)W(z)}\int_{-\infty}^{z}\frac{W(x)}{B(x)}\wrt{x},
\end{equation}
where
\begin{equation}\label{eq:w}
	W(v) = \exp\left(2 \int^{v}\frac{A(u)}{B^2(u)}\wrt{u}\right)
\end{equation}
and where
\begin{align}
\begin{split}\label{eq:general_lif_sol_bounds}
	v_{th}^{eff} &= v_{th} + B(v_{th})\frac{\alpha}{2}k \\
	v_{re}^{eff} &= v_{re} + \frac{B^2(v_{re})}{B(v_{th})}\frac{\alpha}{2}k
\end{split}
\end{align}
are the effective membrane threshold and reset potentials.  Note that $\alpha = -\sqrt{2}\zeta(\frac{1}{2})$ where $\zeta$ is the Riemann zeta function \citep{hagan89halfline}.  Here, \eqref{eq:general_lif_sol} gives the correct two first orders ($0^{\text{th}}$ and $1^{\text{st}}$) in the small $k$ expansion of the firing rate, but also leads to a better approximation of the firing rate in a larger range of values of $k$: it is guaranteed to stay positive at all values of $k$, while $r=\nu_0+k\nu_1$ becomes negative for large $k$.

\subsection*{Comparison of approximation to numeric simulation}

In \figref{comparison} we demonstrate the accuracy of our approximation by simulating a conductance-based LIF neuron described by \eqref{eq:ab} using the following biophysically-realistic parameters: the membrane time constant $\tau_m$ is $\tau_m = C/gl = 37$ ms, the leak conductance is $g_l = 20$ nS, with reversal potentials $v_l = -70$ mV, $v_e = 0$ mV, and $v_i=-80$ mV.  Membrane threshold and reset potentials are $v_{th} = -52 mV$ and $v_{re} = v_l$, respectively \cite{ayaz09modulation}.

\par
A tonic inhibitory conductance of $3g_l$ is included to prevent spiking under noisy excitatory input of zero mean conductance.  Further, we mimic a small amount of balanced synaptic input \cite{ayaz09modulation} by including additional excitatory and inhibitory conductance, leading us to take $\mu_i = 3.3 g_l$ and $\mu_e = (0.1 + \mu)g_l$.  Our qualitative results, however, are insensitive to this parameterization.  \figref{comparison} shows general agreement over a range of parameter values, although the approximation begins to break down as $\tau_s$ increases, consistent with the fact that our analytic formula is valid to first order in $k = \frac{\tau_s}{\tau_m}$.  However, the approximation holds for a realistic membrane constant and input timescales $\tau_s \lessapprox  5$\,ms, similar to estimated decay constants for glutamate and AMPA receptors. \cite{spruston95glutamate}.

\begin{figure}[t]\centering
    \vspace{-5pt}
	\includegraphics[scale=.9]{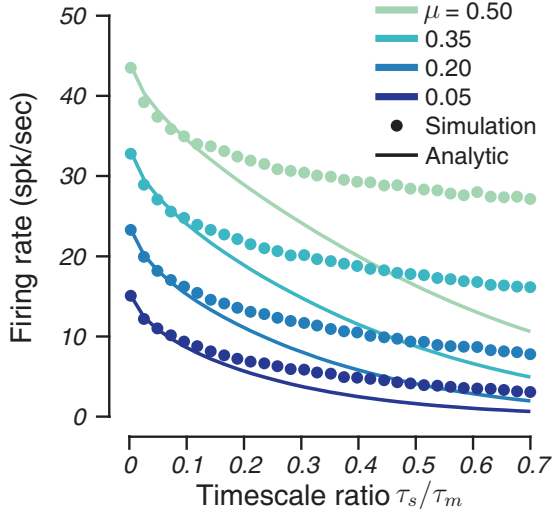}
	\caption{\label{fig:ratio} Comparison of simulation results with the analytic approximation \eqref{eq:general_lif_sol} for \eqref{eq:general_lif} and \eqref{eq:ab} in the case of small balanced synaptic input, \ie $\mu_i = 0.3 g_l$ and $\mu_e = (0.1 + \mu)g_l$ for $\sigma_e = 0.40$.  Note that firing rates exhibit a $\sqrt{\tau_s}^{-1}$ relationship for $\tau_s \ll 1$ across multiple levels of mean excitation $\mu$.}
	\vspace{-5pt}
\end{figure}

The effect of synaptic timescale on the firing rate is examined further in \figref{ratio}.  Here, multiple levels of mean input $\mu$ drive the neuron into a sub- and supra-threshold regime while synaptic fluctuation $\sigma_e = 0.4 g_l$ is fixed.  We note that for small $k$, \ie $\frac{\tau_s}{\tau_m} \lessapprox 0.1$, the approximation closely matches simulation and captures the fact that firing rates are proportional to $\sqrt{\tau_s}^{-1}$.

\section*{Discussion}
In this study we have built upon a previously known approximation to the firing rate of LIF neurons to cover the case of conductance-based input.  Importantly, we find the method to give a good approximation of the firing rate under many biophysically-realistic inputs, providing an analytic tool for studying the response of such neurons.  While we leave a quantitative analysis of approximation error as a topic of future study, the strong qualitative agreement to simulation suggests our derivation to be useful over a range of parameters.  In particular, this work provides an analytic tool for investigating how the statistical properties of input affect a neuron's firing rate, and thus for understanding a neuron's computational properties.

\small
\bibliographystyle{abbrv}
\bibliography{refs}
\end{document}